\documentclass[aps,pra,epsfigure,twocolumn,showpacs]{revtex4}
\usepackage{dcolumn}    
\usepackage{bm} 
\usepackage{graphicx}
\usepackage{amsmath}    
\usepackage{latexsym}
\usepackage{amsfonts}   
\usepackage{amssymb}
\usepackage{array}      
\usepackage{epsfig}

\newcommand{\ket}[1]{\left\vert#1\right\rangle}

\newcommand{\pro}[3]{\left\vert#1\rangle_{#2}\langle#3\right\vert}

\newcommand{\bra}[1]{\left\langle#1\right\vert}
\newcommand{\sand}[3]{\left\langle#1\vert#2\vert#3\right\rangle}

\newcommand{\nbar}{\overline{n}}

\begin{document}
\title{A dissipative scheme to approach the boundary of two-qubit entangled mixed states}
%\title{Approaching the boundary of two-qubit entangled mixed states via dissipation}
%\title{Kylie Minogue (or: I can't get you out of my MEMS..)}
\author{S. Campbell and M. Paternostro}
\affiliation{School of Mathematics and Physics, Queen's University, Belfast BT7 1NN, United Kingdom}

\begin{abstract}
We discuss the generation of states close to the boundary-family of maximally entangled mixed states as defined by the use of concurrence and linear entropy. The coupling of two qubits to a dissipation-affected bosonic mode is able to produce a bipartite state having, for all practical purposes, the entanglement and mixedness properties of one of such boundary states. We thoroughly study the effects that thermal and squeezed character of the bosonic mode have in such a process and we discuss tolerance to qubit phase-damping mechanisms. The non-demanding nature of the scheme makes it realizable in a matter-light based physical set-up, which we address in some details. 
\end{abstract}

\date{\today}
\maketitle

%\section{Introduction}
%\label{Intro}
%There is a growing interest in the field of quantum information theory, and in particular the realms of quantum cryptography and quantum computation. The fundamental theories and algorithms that define these intriguing concepts find their roots firmly in entanglement.

The interplay between entanglement and mixedness has long been recognized as a crucial point to tackle towards a full understanding of the peculiar way correlations of a non-classical nature settle in multipartite configurations of quantum systems~\cite{various}. Despite the impressive efforts produced along these directions and a few important progresses being accomplished, a satisfactory comprehension of such an important topic is still elusive. 
%The concept of entanglement has enjoyed extensive research since its inception but little is still fully understood about it. Only in the last two decades has an acceptable picture for quantifying entanglement been generally adopted, albeit only for the simplest situation of two entangled qubits. Even with much dedicated research to the concept, we still have 
Such incompleteness is well evident in the current unavailabity of a unique and unambiguous entanglement measure for general states involving more than two parties and the striking difficulties related to the ordering of mixed entangled states under different entanglement measures~\cite{ordering}. It is therefore of paramount importance to continue the investigation along these lines. An important contribution to the problem represented by the trade off between entanglement and mixedness has been given by the classification of bipartite states exhibiting the maximum obtainable amount of entanglement for a given degree of mixedness~\cite{MEMS1,MEMS2}. The explicit form of these genuinely interesting states, dubbed maximally entangled mixed states (MEMS's), strongly depends on the chosen measure for entanglement and mixedness.

Devising means of realising these states is of great importance. In a way, as some form of noise will inevitably be present in a physical set-up, the availability of pure states could well be out of question. The interest is thus in achieving the maximum possible entanglement from the mixed-state resource one has to deal with. An efficient state-purification procedure can then be applied to MEMS's as described in Ref.~\cite{ogden}. Work at all levels has been performed on the generation of MEMS's. Linear-optics settings involving parametric down-conversion processes have been used in order to experimentally explore quite a substantial region of the physically allowed entanglement-mixedness space, up to the MEMS boundary~\cite{kwiat,demartini}. Theoretical proposals have been put forward for the {\it navigation} in the plane of entangled mixed states~\cite{clark,paternostro}, involving multi-level atom-like objects either interacting with structured environments or following properly arranged unitary evolutions. Interestingly, Cho and McKenzie have proposed a scheme for the generation of bipartite Werner states in a two-impurity Kondo model via the well-known Ruderman-Kittel-Kasuya-Yosida interaction~\cite{mckenzie}. This is interesting as, under proper choices of entanglement and mixedness quantificators, Werner states {\it are} MEMS's~\cite{MEMS1}. Here, we address a simple scheme that allows a system of two qubits to approach the boundary of physically allowed entangled mixed states. The protocol is based on biased spin-boson interaction under the influences of dissipation and phase-damping. Besides its simplicity, the scheme addressed here shows that dissipation is able to coherently lead the qubit system to a partially entangled mixed state which, nevertheless, is interestingly close to the MEMS boundary. We show that the availability of a {\it cold enough} environment, together with an asymmetric preparation of the qubit system is all we need in order to approach such boundary curve. The scheme is highly realistic, as it estimates and includes the effects of the most relevant entanglement-spoiling mechanisms and, as we discuss, holds the promises for a prompt experimental realization in set-ups of cavity as well as circuit-quantum electrodynamics (QED). 

The remainder of this paper is organized as follows. Sec.~\ref{MEMS} briefly introduces MEMS's and their properties, besides discussing the main technical tools used throughout the paper. In order to fix the ideas, Sec.~\ref{unitary} gives an account of the unitary version of the scheme discussed here while the main part of our analysis is presented in Sec.~\ref{Diss}. There, we address the reduced dissipative dynamics undergone by the qubit system and identify a state of closest proximity to an element of the MEMS family. Some technical details, unnecessary to the comprehension of our main results, are presented in Appendix A. In Secs.~\ref{squeezed} and \ref{phase} we give account of how a structured environment as well as the introduction of phase-damping mechanism would affect our findings. Sec.~\ref{practical} describes in some details an experimental set-up that has the necessary features for the implementation of the physical mechanisms assessed here. Finally, Sec.~\ref{remarks} provides a summary of our findings and conclusions.

\section{Introduction to Maximally Entangled Mixed States}
\label{MEMS}

%An important class of states in quantum information theory (QIT) are the classes of Maximally Entangled Mixed States (MEMS)~\cite{MEMS1}. These states exhibit the maximum obtainable entanglement for a given mixedness and so are the epitome of experimentally realisable states.

Here, we briefly remind the basic properties of MEMS's and their parameterization for a specific choice of entanglement and mixedness measure. As we mentioned, a long-standing argument regards state-ordering induced by entanglement measures~\cite{MEMS2}. It has been seen that for different entanglement measures there are different possible parameterisations of MEMS's. To date, bipartite MEMS's have been found to be described by a one-parameter family of density matrices~\cite{MEMS1}. For the purposes of this paper, we shall restrict ourselves to the entanglement measure given by concurrence~\cite{CONC}, which is defined in relation to the entanglement of formation. The latter quantifies the number of Bell states required to prepare a given state. Entanglement of formation depends monotonically on concurrence which, for a two-qubit (pure or mixed) state $\rho$ can be defined as~\cite{NIELSEN,CONC}
\begin{equation}
C= \max [0,\sqrt{\lambda_1}-\sqrt{\lambda_2}-\sqrt{\lambda_3}-\sqrt{\lambda_4} ]
\end{equation}
where $\lambda_1\ge\lambda_j~(j=2,3,4)$ are the eigenvalues of $\rho(\sigma_y \otimes \sigma_y) \rho^* (\sigma_y \otimes \sigma_y)$ and $\sigma_y$ is the $y-$Pauli spin operator.  We shall be measuring mixedness of a state via linear entropy (see for example Bose and Vedral in~\cite{various})
\begin{equation}
S=\frac{4}{3}[1-{\rm Tr}(\rho^2)].
\end{equation}
The necessary parameterisation for MEMS with regards to these measures was provided in~\cite{MEMS1}, and identifies two subclasses of density matrices
\begin{equation}
\rho_1 = 
\left(
\begin{array}{llll}
 \frac{r}{2} & 0 & 0 & \frac{r}{2} \\
 0 & 1-r & 0 & 0 \\
 0 & 0 & 0 & 0 \\
 \frac{r}{2} & 0 & 0 & \frac{r}{2}
\end{array}
\right),~~~\rho_2 =
\left(
\begin{array}{llll}
 \frac{1}{3} & 0 & 0 & \frac{r}{2} \\
 0 & \frac{1}{3} & 0 & 0 \\
 0 & 0 & 0 & 0 \\
 \frac{r}{2} & 0 & 0 & \frac{1}{3}
\end{array}
\right),
\end{equation}
where $\rho_1$ ($\rho_2$) holds for $r\in[2/3,1]$ ($r\in[0,2/3]$). In a $C$ versus $S$ plane, these states lie on the so-called MEMS boundary curve shown in Fig. 1 {\bf (b)}, where the upper (lower) portion of the solid curve, {\it i.e.} the part corresponding to $C\ge{2/3}$ ($C\le{2/3}$), is for $\rho_1$ ($\rho_2$).\\

\begin{figure}[t]
{\bf (a)}\hskip3cm{\bf (b)}
\psfig{figure=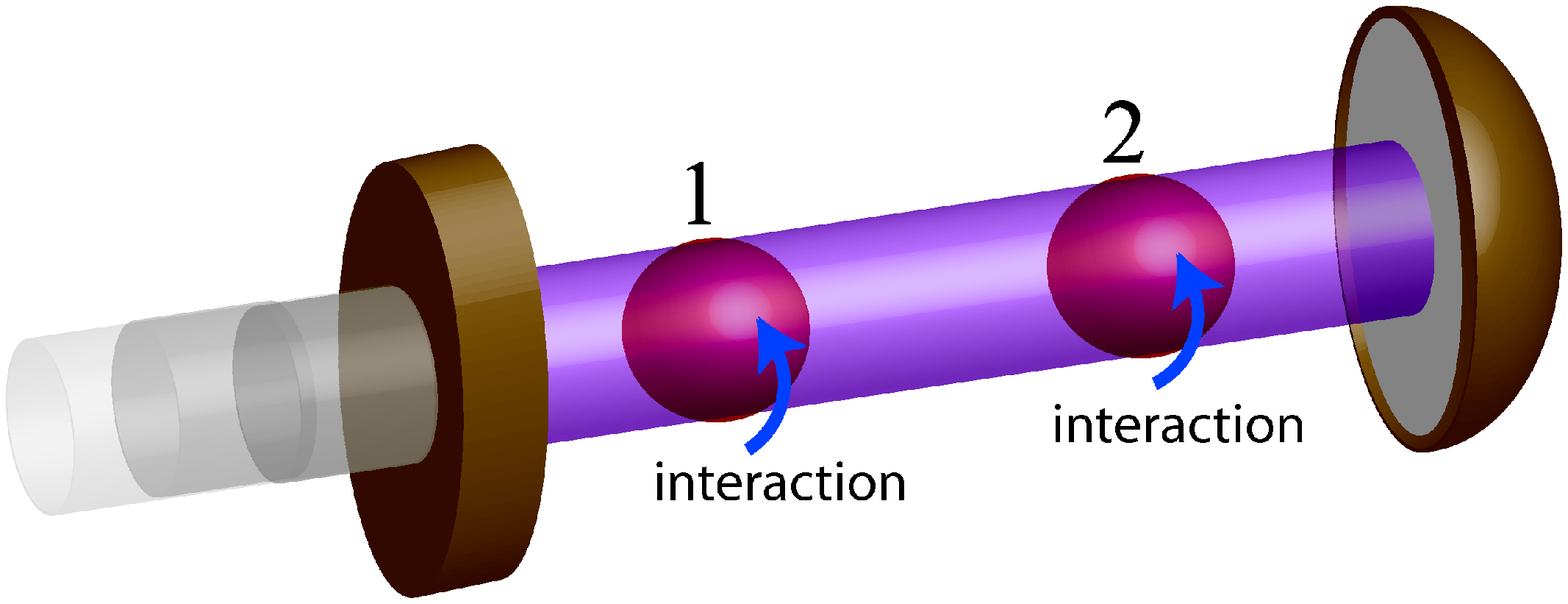,width=4.8cm,height=2.0cm}~~\psfig{figure=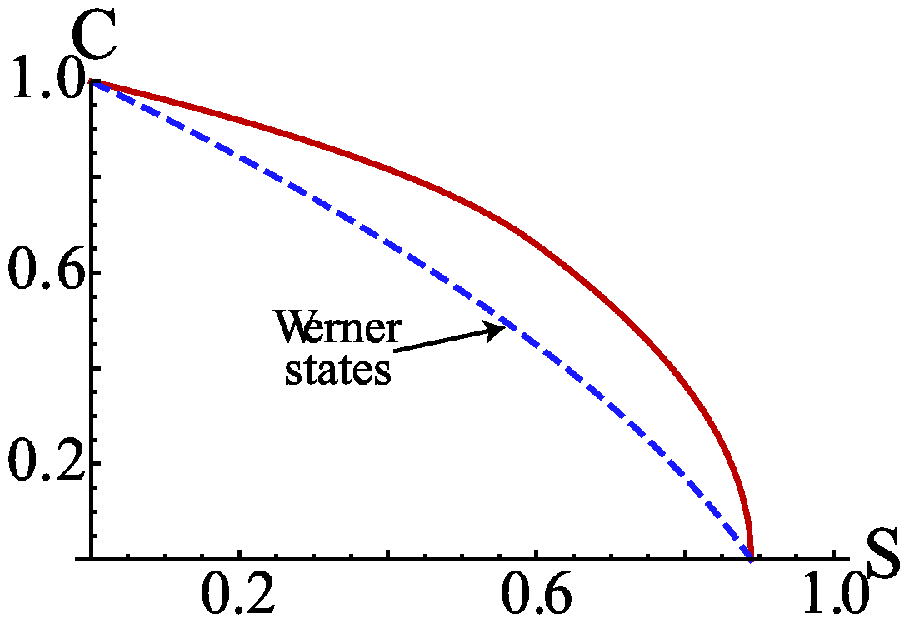,width=3.5cm,height=2.5cm}
\caption{(Color online). {\bf (a)} Sketch of the discussed {\it thought experiment}. A cavity accommodates two two-level systems (labelled $1$ and $2$), asymmetrically coupled to the cavity field. The latter leaks out of the cavity due to a finite quality factor. {\bf (b)} Concurrence versus linear entropy MEMS boundary (full line). For comparison, the dashed lines shows the curve corresponding to the two-qubit Werner state $\rho_{W}=p\ket{\Phi}\bra{\Phi}+(1-p)\openone/4$ with $p\in[0,1]$ and $\ket{\Phi}=(1/\sqrt{2})(\ket{00}+\ket{11})_{12}$.}
\label{fig1}
\end{figure}

\section{Approaching MEMS boundary}
\subsection{Unitary case}
\label{unitary}

We start introducing the coupling model considered throughout our work, in the idealized situation of a perfectly unitary evolution. 
%Two qubits, labelled from now one $1$ and $2$, are asymmetrically coupled  to a common bosonic mode $b$ by means of  bilocal resonant Jaynes-Cummings like interactions~\cite{JC}. 
Usually, problems involving the interaction of two-level systems with a boson can be modelled using an effective model where a dipole-like spin operator (proportional to the $\hat{\sigma}_x$ Pauli operator) couples to the electric (or the magnetic) part of a field. This is the case for neutral atoms or quantum dots coupled to optical fields. However, this description holds also for a system consisting of a Cooper-pair box [in a superconducting-quantum-interference device (SQUID) configuration and in the charge regime~\cite{schon}] integrated into a planar stripline resonator~\cite{schoelkopf}, a setting generally referred to as {\it circuit-QED}. At the charge degeneracy point, an effective dipole moment operator for the SQUID can be written, whose amplitude is proportional to the excess charge in the SQUID island~\cite{schoelkopf}. Here, in order to fix the ideas and introduce the general formalism employed throughout our study, we use language and terminology typical of cavity-QED and we refer explicitly to a scheme of atomic qubits interacting with the field of an optical cavity. In Sec.~\ref{practical} we assess the details of possible experimental implementations.

\begin{figure}[b]
{\bf (a)}\hskip4.5cm{\bf (b)}
\psfig{figure=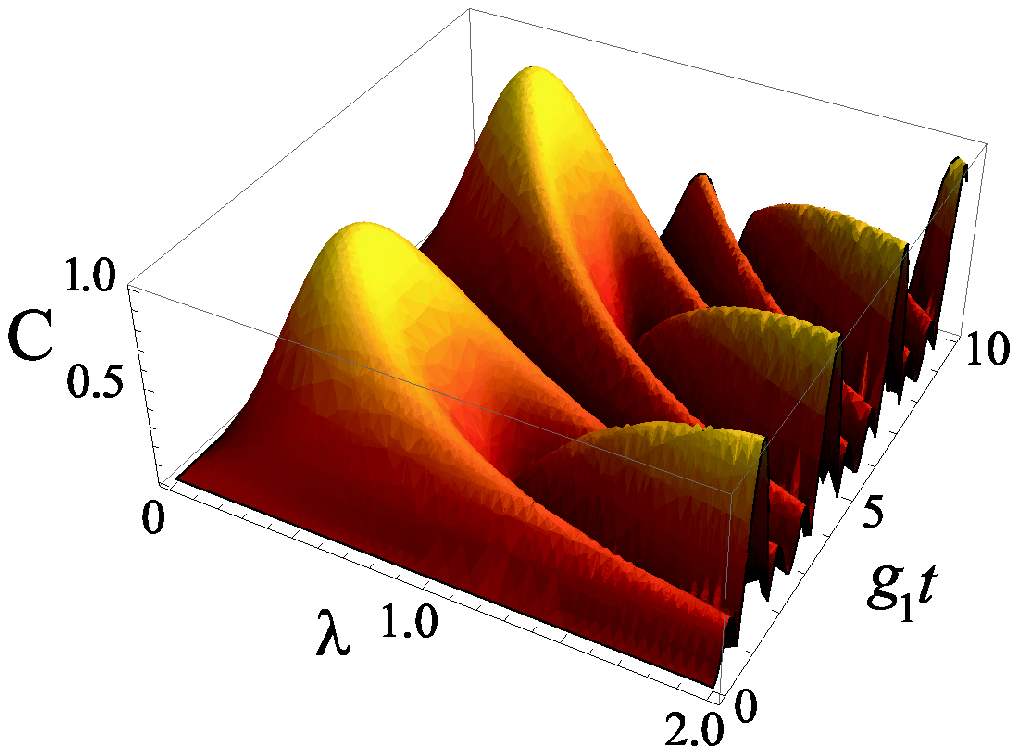,width=4.5cm,height=3.5cm}~~\psfig{figure=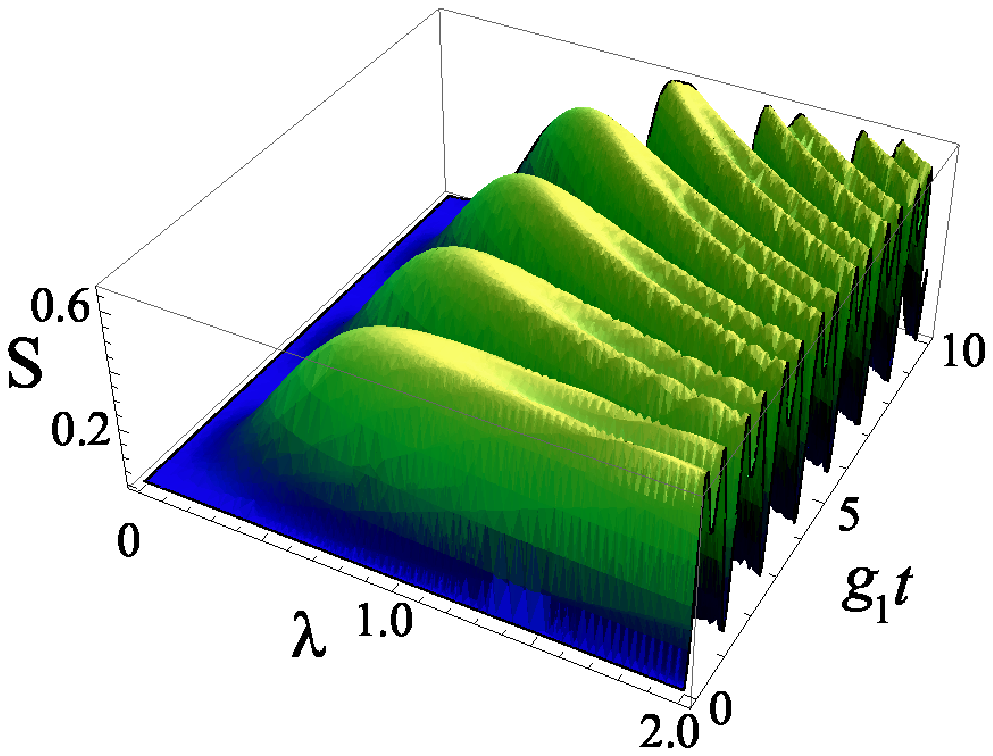,width=4.5cm,height=3.5cm}
\caption{(Color online). {\bf (a)} Concurrence of the produced $\varrho_u(\tau)$ against the rescaled interaction time $\tau=g_{1}t$ and the coupling ratio $\lambda$. {\bf (b)} Mixedness of the state against $g_{1}t$ and $\lambda$.}
\label{Unitary1}
\end{figure}

We consider two qubits with ground and excited states $\ket{0}$ and $\ket{1}$, respectively. They are allocated into a single-mode cavity and have the same transition frequency $\omega_o$, resonant with the frequency $\omega_f\simeq{\omega}_o$ of the cavity field. This is described by the bosonic annihilation (creation) operator $\hat{a}$ ($\hat{a}^{\dag}$). Within the dipole-coupling interaction assumed here and using the rotating-wave approximation~\cite{gerryknight} in interaction picture with respect to the free Hamiltonian of the system, the coupling reduces to (we assume units such that $\hbar=1$ throughout the paper)
\begin{equation}
\label{coupling}
\hat{\cal H}_{I}=\sum^2_{j=1}g_j(\hat{\sigma}^{-}_j\hat{a}^\dag+\hat{\sigma}^{+}_j\hat{a}),
\end{equation}
where $\hat{\sigma}^+_j=(\hat{\sigma}^-_j)^\dag=\pro{1}{j}{0}$ is the qubit raising operator and $g_j$ is the coupling strength of qubit $j$ with the field. Fig.~\ref{fig1} {\bf (a)} shows a sketch of the idealized situation considered here. We assume that the qubit system is initally prepared in state $\ket{\psi(0)}_{12}=\ket{01}_{12}$, while the cavity field is in $\ket{0}_b$. Eq.~(\ref{coupling}) commutes with the operator counting the total number of excitations within the system, so that the corresponding dynamics can be studied within finite-dimension subspaces. By considering the single-excitation subspace (consistently with our initial-state assumption), it is straighforward to see that solving the time-dependent Schr\"odinger equation is equivalent to finding the solution to the following set of coupled linear differential equations
\begin{equation}
\label{set}
\begin{split}
i\partial_{t}\alpha(t)&=g_1\chi(t),~~i\partial_t\beta(t)=g_2\chi(t),\\
i\partial_t\chi(t)&=g_1\alpha(t)+g_2\beta(t),
\end{split}
\end{equation}
where we have used the decomposition $\alpha(t)\ket{100}_{12b}+\beta(t)\ket{010}_{12b}+\chi(t)\ket{001}_{12b}$ for the state of the whole system at time $t$ (with $|\alpha|^2+|\beta|^2+|\chi|^2=1$). By introducing the coupling ratio $\lambda=g_2/g_1$ and the dimensionless interaction time $\tau=g_1{t}$, the density matrix of qubits $1$ and $2$ obtained by tracing out the field's state is found to be
\begin{equation}
\label{stateunitary}
\varrho_{u}(\tau)=
\begin{pmatrix}
\chi^2(\tau)&0&0&0\\
0&\beta^2(\tau)&\beta(\tau)\alpha(\tau)&0\\
0&\beta(\tau)\alpha(\tau)&\alpha^2(\tau)&0\\
0&0&0&0
\end{pmatrix}
\end{equation}
with $\chi(\tau)=\lambda\sin(\tau\sqrt{1+\lambda^2})/\sqrt{1+\lambda^2},\beta(\tau)=[1+\lambda^2\cos(\tau\sqrt{1+\lambda^2})]/(1+\lambda^2)$ and $\alpha(\tau)=-\lambda(1-\cos(\tau\sqrt{1+\lambda^2})(1+\lambda^2)$. It is worth noting that in tracing out the field the observed dynamics are no longer evolving unitarily with respect to $\tau$. Concurrence and linear entropy of $\varrho_u(\tau)$ can be easily calculated. Their behavior is shown in Figs.~\ref{Unitary1} against $\tau$ and $\lambda$.
By inspection, it is clear that concurrence and mixedness are maximized at small, non-unit values of the coupling ratio, when a substantial amount of entanglement can be found in qubit states with $S\ll{1}$. Of course, mixedness of the state here arise in virtue of the loss of information over the field state. Shown in a $C-S$ plane, these features make it evident that density matrices belonging to the boundary of physically meaningful states can be generated for a proper choice of $\tau$ and $\lambda$.
\begin{figure}[t]
\psfig{figure=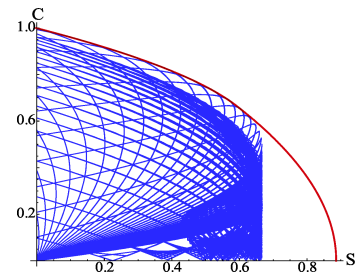,width=4.5cm,height=3.5cm}
\caption{(Color online). Concurrence $C$ versus linear entropy $S$ in $\rho_{u}(\tau)$ and comparison with the MEMS boundary. Each continuous (blue) curve corresponds to an open trajectory of the two-qubit state having $\lambda\in[0,2]$ (which increases at steps of $0.05$) and the curvilinear abscissa is $\tau\in[0,40]$.}
\label{Unitary2}
\end{figure}
Interestingly, despite the rather large temporal range we have considered ($\tau\in[0,40]$), the elements of the MEMS family $\rho_2$ cannot be produced by this scheme, which priviledges highly-entangled states (up to $C=1$) of various mixedness. Differently from what was found in Ref.~\cite{clark}, our states do not ``track'' the behavior of the MEMS boundary but follow a dynamical {\it touch-and-go} pattern with respect to $\rho_1$. The similarity between $\varrho_{u}(\tau)$ and $\rho_1$ (for certain values of $r,\lambda$ and $\tau$) can be clearly understood by relying on the spectral decomposition of these states: at the ``touching'' point between the open trajectories associated with two-qubits states shown in Fig.~\ref{Unitary2} and the MEMS curve, the eigenvalues (eigenvectors) of $\varrho_{u}$ are identical (locally equivalent) to those of $\rho_1$~\cite{example}. This is never the case for states $\varrho_{u}(\tau)$ and $\rho_2$.

\subsection{Approaching MEMS boundary: Open-system dynamics}
\label{Diss}

Although promising because of the possibility of generating a vast range of boundary states, the scheme described so far may be far from being realistic, in some set-ups. A unitary description is hardly retained for the whole range of time necessary in the MEMS-approaching mechanism when systems of quantum dots (neutral atoms) embedded in a cavity are considered, for instance. Any realistic physical set-up 
will imply the consideration of a finite rate of amplitude (phase) damping affecting the system at hand. In particular, for the specific instance considered in this work, energy leakage outside the cavity due to finite resonator quality factor should be quantitatively included in our calculations. This is extremely important within the context of our investigation, especially in virtue of the special role played by the excitation-conservation rules highlighted above. The introduction of an energy-dissipation mechanism, as it is the case with a leaky cavity, breaks such conservation law forcing us to study the dynamics of the system in the whole Hilbert space, in principle. The effects of such differences should be carefully quantified.

This is precisely what we do here, where we replace the Schr\"odinger equation at the basis of our study so far, with a dissipative master equation for the state of the qubits-field system $\varrho_c(t)$ reading (unless otherwise specified, we use again the notation involving $g_{1,2}$ and $t$)
\begin{equation}
\label{ME}
\partial_t\varrho_c(t)=-i[\hat{\cal H}_I,\varrho_c(t)]+\hat{\cal L}[\varrho_c(t)]\equiv(\hat{\cal L}_u+\hat{\cal L})[\varrho_c(t)]
\end{equation}
with the Liouville superoperator 
\begin{equation}
\begin{split}
\hat{\cal L}[\varrho_c(t)]&=\gamma(\nbar+1)(2\hat{a}\varrho_c(t)\hat{a}^\dag-\{\hat{a}^\dag\hat{a},\varrho_c(t)\})\\
&+\gamma\nbar(2\hat{a}^\dag\varrho_c(t)\hat{a}-\{\hat{a}\hat{a}^\dag,\varrho_c(t)\}).
\end{split}
\end{equation}
Here, $2\gamma$ is the energy dissipation rate from the cavity and $\nbar$ is the mean thermal occupation number of the bath with which the field is at equilibrium \cite{notatemp}. Throughout our treatment, we assume the {\it bad cavity limit} $\gamma\gg{g_{1,2}}$. This implies that the cavity field response to the bath is much faster than that associated with its interaction with the qubits. In turn, this means that we can neglect the backaction of the photons emitted by the qubits therefore validating the Born approximation. Moreover, the bad-cavity limit also allows for Markovian treatment of the open dynamics as it is equivalent to a short qubit memory time. In analogy to what has been done in the unitary case, we want to trace out the field degrees of freedom, which can be systematically done by using projection-operator techniques~\cite{gardiner}. However, in order to bypass the technicalities involved in such an approach, here we follow a simpler and yet rigorous way which leads to the same results as the projection-operator technique~\cite{agarwalpuri}. 

In the bad cavity limit, it is reasonable to assume that the cavity field remains in the thermal steady state determined by $\hat{\cal L}$, which we label $\rho^{ss}$. We remind that within the bad cavity limit we are necessarily assuming weak coupling of the qubits with the field with respect to the field dissipation rate. In what follows, the validity of our results holds in this limit. Upon trace over the field state we get the following formal solution for the evolution of the qubits alone
\begin{equation}
\label{formalsol}
\partial_t\varrho(t)=\int^t_0{\rm Tr}_{\text{field}}\{\hat{\cal L}_ue^{\hat{\cal L}(t-t')}\hat{\cal L}_u[\varrho(t')\otimes\rho^{ss}]\}dt'.
\end{equation}
%definition, we have that $\hat{\cal L}[\rho^{ss}_b]=0$, which helps us in finding that
Moreover, we can easily find that
\begin{equation}
\label{relations}
\begin{split}
\hat{\cal L}_u(\hat{a}\varrho_c)&=\gamma(2\nbar+1)\hat{a}\varrho_c-2\gamma\nbar\varrho_c\hat{a},\\
\hat{\cal L}_u(\varrho_c\hat{a})&=2\gamma(\nbar+1)\hat{a}\varrho_c-\gamma(2\nbar+1)\varrho_c\hat{a}
\end{split}
\end{equation} 
with $\varrho_c=\varrho\otimes\rho^{ss}$. Upon iteration of these relations and explicit evaluation of the commutators involved in Eq.~(\ref{formalsol}), a lengthy but straightforward calculation leads to the reduced master equation for systems $1$ and $2$
\begin{equation}
\label{reduced}
\begin{split}
\partial_t\varrho(t)&=\sum^2_{j=1}\frac{g^2_j}{\gamma}\{(\nbar+1)(2\hat{\sigma}^-_j\varrho(t)\hat{\sigma}^+_j-\{\hat{\sigma}^j_+\hat{\sigma}^j_-,\varrho(t)\})\\
&+\nbar(2\hat{\sigma}^+_j\varrho(t)\hat{\sigma}^-_j-\{\hat{\sigma}^j_-\hat{\sigma}^j_+,\varrho(t)\})\}\\
&+\frac{g_1g_2}{\gamma}\sum^2_{j\neq{k}=1}\{(\nbar+1)(2\hat{\sigma}^-_j\varrho(t)\hat{\sigma}^+_k-\{\hat{\sigma}^+_j\hat{\sigma}^-_k,\varrho(t)\})\\
&+\nbar(2\hat{\sigma}^+_j\varrho(t)\hat{\sigma}^-_k-\{\hat{\sigma}^-_j\hat{\sigma}^+_k,\varrho(t)\})\}.
\end{split}
\end{equation}
Eq.~(\ref{reduced}) is the starting point of our analysis. It can be solved by projecting it onto states of the two-qubit computational basis $\{\ket{00},\ket{01},\ket{10},\ket{11}\}_{12}$, in a way so as to study the Bloch-like differential equations for the density matrix elements $\varrho_{ijkl}={}_{12}\!\sand{ij}{\varrho}{kl}_{12}$ ($i,j,k,l=0,1$). The explicit form of such equations is given Appendix A. Here, we discuss the results achieved by solving them.

\begin{figure}[b]
\psfig{figure=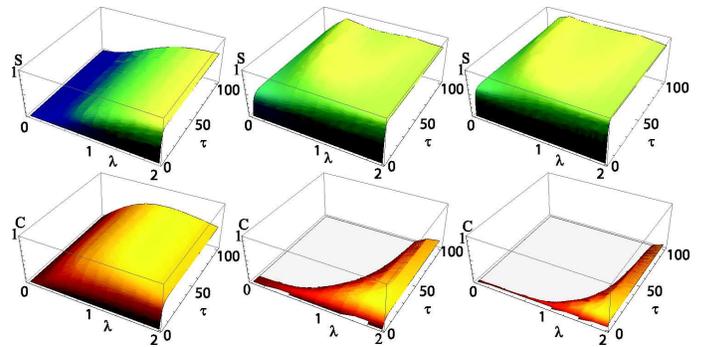,width=9cm,height=4.5cm}
\caption{(Color online). We show the behavior of linear entropy $S$ (first row of plots) and concurrence $C$ (second row) against $\lambda$ and $\tau=g_1t$ for $\nbar=0,0.4$ and $0.8$, in going from leftmost to rightmost plot in each row, respectively.}
\label{Insieme}
\end{figure}

Our initial state is again $\ket{01}_{12}$ and the density matrix is found to have the general form
\begin{equation}
\label{rhogen}
\varrho(t)=
\left(
\begin{array}{llll}
 \varrho_{0000}(t) & 0 & 0 & 0 \\
 0 & \varrho_{0101}(t) & \varrho_{0110}(t)
   & 0 \\
 0 & \varrho_{0110}(t) & \varrho_{1010}(t)
   & 0 \\
 0 & 0 & 0 & \varrho_{1111}(t)
\end{array}
\right).
\end{equation}
The presence of the non-zero $\rho_{1111}(t)$ element is a first significant difference with respect to Eq.~(\ref{stateunitary}). In fact, while it is easy to see that in the unitary case concurrence was simply determined by the off-diagonal density matrix elements, here $C$ depends critically on a delicate trade off between populations and coherences. Explicitly
\begin{equation}
C= -2\varrho_{0110}(t)-2 \sqrt{\varrho_{0000}(t)\varrho_{1111}(t)},
\end{equation}
where we have used the fact that $\varrho_{0110}(t)<0$ for any choice of the parameters involved and at any time.
%\begin{equation}
%S_L (\rho)  =  \frac{4}{3} \left(1-\text{$\rho_{0000}$}^2-\text{$\rho_{0101}$}^2-2\text{$\rho_{0110}$}^2-\text{$\rho_{1010}$}^2-\text{$\rho_{1111}$}^2 \right)
%\end{equation}
We now re-introduce the coupling ratio $\lambda$ and the dimensionless time $\tau$ and analyze the behavior of $C$ and $S$ corresponding to state (\ref{rhogen}). This is done in Figs.~\ref{Insieme}, where $\lambda\in[0,2]$ with $\tau\in[0,100]$ are taken and concurrence (linear entropy) is studied for specific values of $\nbar$.

\begin{figure}[t]
{\bf (a)}\hskip4cm{\bf (b)}
\psfig{figure=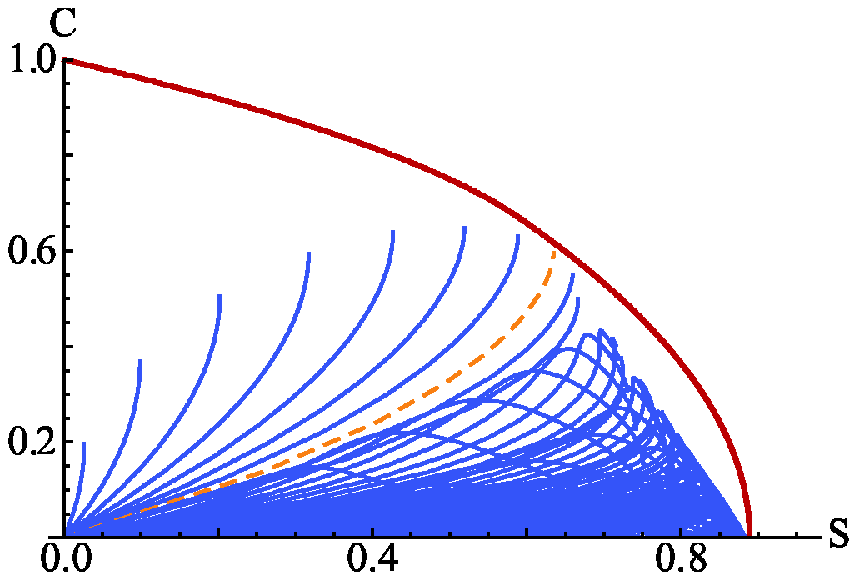,width=4.2cm,height=3.2cm}~~\psfig{figure=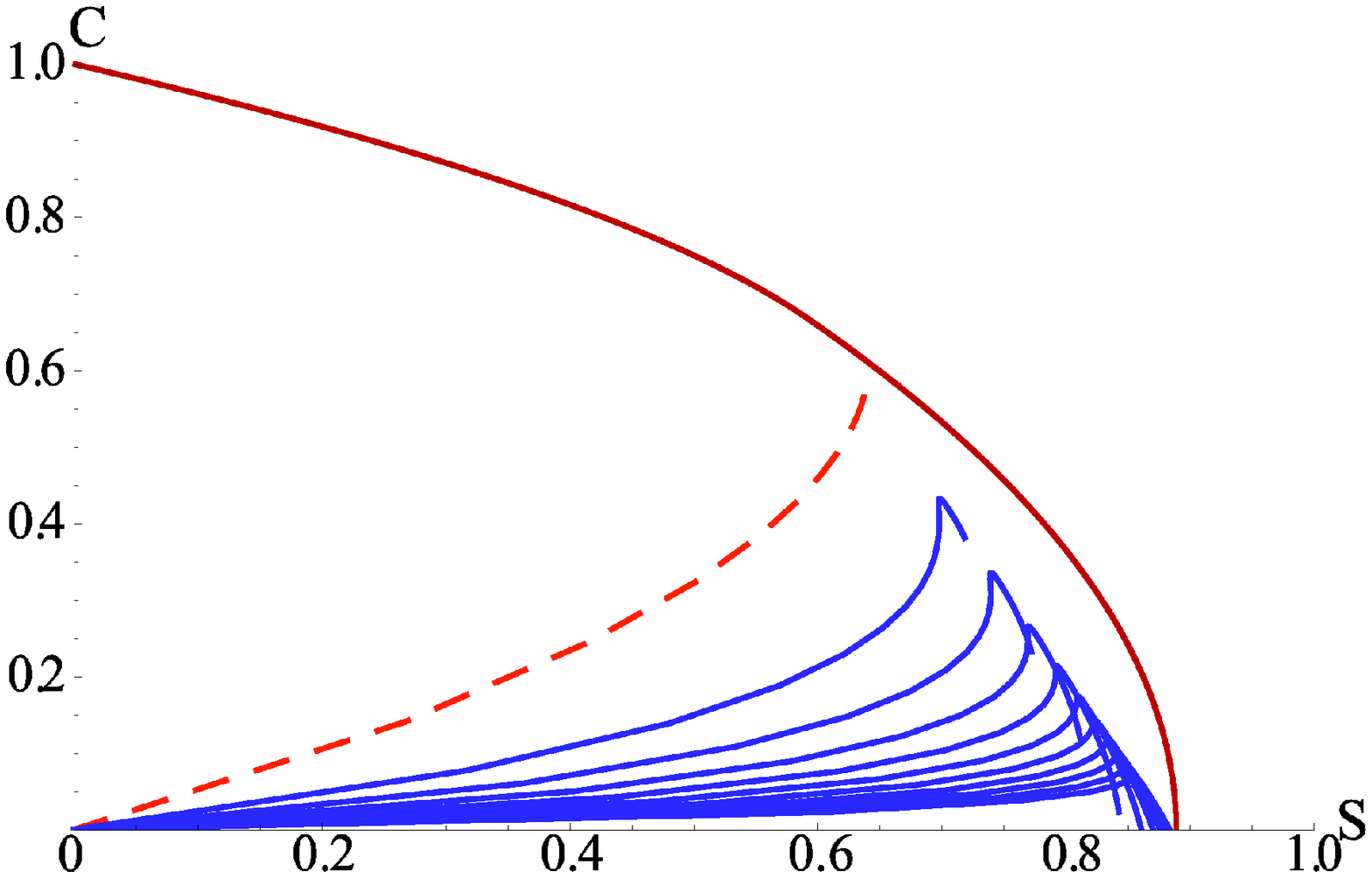,width=4.2cm,height=3.2cm}
\caption{(Color online). $C-S$ plot for $\varrho(\tau)$ as determined by a dissipative dynamics with $\nbar\ge{0}$ and $\gamma/g_1=10$. The dashed curve embodies the open trajectory for the optimal state corresponding to $\nbar=0$ and $\lambda=\lambda_{opt}$. {\bf (b)} Decay of entanglement and increase of mixedness in the optimal state for $\nbar$ that goes from $0$ to $1$ (in steps of $0.1$, from top to bottom open trajectory).}
\label{fig2}
\end{figure}

Immediately, one recognizes that $C$ is maximized at $\lambda<1$, although the actual value is sensitive to $\nbar$. Quite expectedly, in virtue of the features characterizing the unitary case, $\nbar=0$ corresponds to the maximum (minimum) of $C$ ($S$) evaluated over the stationary state of the qubit system. This is reached already for $\tau\simeq{30}$. A numerical inspection reveals that $\lambda_{opt}\sim0.8$ corresponds to the largest possible value of concurrence of a state having $S<0.7$. In the $C-S$ plane, these results are summarized in Fig.~\ref{fig2} {\bf (a)}, where curves associated with increasing values of $\lambda$ and $\nbar$ are compared to the MEMS boundary. The (dashed) curve corresponding to $\lambda_{opt}$ and $\nbar=0$ is clearly highlighted. The tip of this curve corresponds to a state which is extremely close to the MEMS boundary. However, a few remarks are in order: differently from the unitary case, the dissipative scheme is able to approach family $\rho_2$, which belongs to the lower part of the MEMS boundary curve. Dissipation thus depletes the general properties of the generated states. Although very close to the boundary, $\varrho(\tau)$ never touches it. However, the scheme does not require any time control. In fact, the state of closest distance from MEMS is achieved as a steady state: for $\tau>30$ the properties of $\varrho$ do not change and the curves in Fig.~\ref{fig2} {\bf (a)} do not fold back as time grows.
%\begin{figure}[ht]
%\psfig{figure=ThermalEffectsonOPtN.eps,width=5.5cm,height=4.5cm}
%\caption{Thermal Effects on Optimal Solution}
%\label{fig3a}
%\end{figure}

\begin{figure}[b]
{\bf (a)}\hskip4cm{\bf (b)}
\psfig{figure=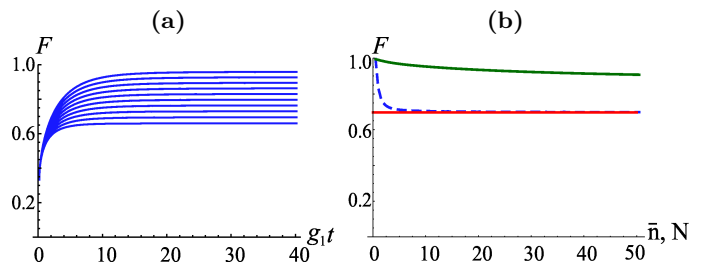,width=9cm,height=3.cm}
\caption{(Color online). {\bf (a)} State fidelity $F(\varrho,\rho_2)$ against the 
dimensionless interaction time $\tau=g_1t$ for $\lambda=\lambda_{opt}$ and 
$r\in[0,2/3]$ increasing at steps of $2/30$ (going from bottom- to top-curve). 
{\bf (b)} The dashed (solid) curve shows $F(\varrho,\rho_2)$ 
against the parameter $\nbar$ ($N$) characterizing the thermal (squeezed) 
bath for $r\simeq{2/3}$, $g_1{t}=100$ and $\lambda=\lambda_{opt}$. The 
horizontal line shows the fidelity between $\rho_{2}$ (with the above 
choice for $r$) and the maximally mixed state $\openone/4$.}
\label{fig3}
\end{figure}

The analytic form of the optimal state at any $\tau$ can be found from the explicit solution of Eqs.~(A-1) to (A-9).
\begin{equation}
\label{correction}
\begin{split}
&\varrho_{0000}(\tau)\!=\!\frac{\lambda^2\left(1-e^{-\frac{2G\tau}{\gamma }}\right)}{G^2},\varrho_{0101}(\tau)\!=\!\frac{\left(1+e^{\frac{-G\tau}{\gamma }}\lambda^2\right){}^2}{G^4},\\
&\varrho_{1010}(\tau)\!=\!\frac{\lambda^2(1-e^{\frac{-G\tau}{\gamma }})^2}{G^4},\varrho_{0110}(\tau)\!=\!-\sqrt{\varrho_{1010}(\tau)\varrho_{0101}(\tau)}
\end{split}
\end{equation}
%\\
%and
%\begin{equation}
%\rho_{0110}=-\frac { g_1 g_2\left(1-e^{-\frac{
%  Gt}{\gamma }}\right)\left(g_1^2+ g_2^2e^{-\frac{Gt}{\gamma
%   }}\right)}{G{}^4}
%\end{equation}
with $\varrho_{1111}(\tau)=0$ and $G=\sqrt{g_1^2+g_2^2}=g_1\sqrt{1+\lambda^2}$. It is interesting to notice that such an optimal state enjoys the same features as $\varrho_u(\tau)$, {\it i.e.} the absence of populations of the $\ket{11}_{12}$ state, which effectively leaves the two-qubit state in a one-excitation subspace. As for the unitary case, $C$ is determined simply by the off-diagonal elements of the density matrix. We can quantify exactly how close we get to the $\rho_2$ class by using state fidelity $F(\rho_A,\rho_B)$~\cite{NIELSEN}, which gives an estimate of the similarity between two density matrices $\rho_A$ and $\rho_B$. For identical (orthogonal) states, $F=1$ ($F=0$). Here, we use the ``amplitude'' version of $F$ defined by $F(\rho_A, \rho_B) = \text{Tr}\sqrt{\sqrt{\rho_A}\rho_B \sqrt{\rho_A} }$, which is not too sensitive to very small changes in the optimal set of parameters. By comparing $\rho_2$ and the density matrix having elements given by Eqs.~(\ref{correction}), we immediately recognize that $F(\varrho,\rho_2)$ cannot be close to $1$ as the position of the coherence terms in the two density matrices do not correspond. Moreover, while $r\in[0,2/3]$ in $\rho_2$ guarantees positive coherences, we have that $\varrho_{0110}(\tau)\le0~\forall{\tau}$. And yet, the properties of the two states are rather close, as we have already commmented. However, it is straigthforward to see that, by means of a bit and phase flip on qubit $2$, we get $\varrho\simeq{\rho_2}$. Needless to say, such local unitaries leave $C$ or $S$ invariant. The behavior of $F(\varrho,\rho_2)$ against $\tau$ is shown in Fig.~\ref{fig3} {\bf (a)} for increasing values of $r$ in the range valid for $\rho_2$. At $r\sim{2/3}$, $\gamma=10g_1$ and $\lambda=0.8$ we find $F(\varrho,\rho_2)>99.4\%$. This choice of parameters define the state (for $\tau\ge40$)
\begin{equation}
\varrho=
\begin{pmatrix}
 0.398 & 0 & 0 & 0 \\
 0 & 0.362 & -0.295 & 0 \\
 0 & -0.295 & 0.24 & 0 \\
 0 & 0 & 0 & 0
\end{pmatrix}
\end{equation}
which has concurrence and linear entropy values of $C=0.589,~S=0.639$. Fig.~\ref{fig2} {\bf (b)} and Fig.~\ref{fig3} {\bf (b)} show the effects induced of the thermal nature of the field on such an optimal state. In particular, state fidelity decreases quite rapidly towards $\sim0.7$. Despite such a relatively large asymptotic value, this implies that the corresponding state is quite far from a MEMS, as seen in Fig.~\ref{fig2} {\bf (b)} too. In fact, $F=0.7$ is the state fidelity between the maximally mixed state $\openone/4$ and $\rho_2$.

\subsection{Irrelevance of the use of squeezed light
 for MEMS boundary approach}
\label{squeezed}

Based on the results of Refs.~\cite{clark,paternostro}, where physical or effective squeezed fields where beneficial to the task of MEMS's generation, one may now wonder whether coupling the cavity field to a squeezed bath can improve the performances of the protocol addressed here. In this section we briefly show that, for the specific instance here studied, this is not the case at all. 

In order to provide a quantitative study to this situation, we have derived the reduced master equation for two qubits immersed in the field of a cavity coupled to a broadband squeezed-vacuum characterized by the parameters $N$ and $M\le\sqrt{N(N+1)}$ related to the squeezing parameter $s$ (the equality sign holding for ideal squeezing, in which case we have $N=\sinh^2{s}$). This can efficiently be done by writing Eq.~(\ref{reduced}) for $\nbar=0$ and replacing raising and lowering qubit operators with their ``squeezed'' versions $\hat{\Sigma}^+_j=\sqrt{N}\hat{\sigma}^+_j+\sqrt{M/N}\hat{\sigma}^-_j$ and $\hat{\Sigma}^-_j=\sqrt{M/N}\hat{\sigma}^+_j+\sqrt{N}\hat{\sigma}^-_j$. The resulting master equation, depending explicitly on $N,M$ and the effective coupling rates $g^2_j/\gamma$ and $g_1g_2/\gamma$, is identical to the one obtained adapting the approach described in~\cite{cirac} which passes through the use of a squeezed and dissipative picture. For $M=0$, the thermal cavity-field case is found back, while the case of both $M=N=0$ corresponds to the optimal vacuum-field situation that has been extensively discussed above. Bloch-like equations can be derived and an analysis analogous to the one described in Sec.~\ref{Diss} performed. However, one recognizes that the introduction of a squeezed bath in the system considered in our study does not help the MEMS-approaching task. In fact, the situation described in Figs.~\ref{Insieme} seems to optimize both the entanglement and mixedness within the two-qubit system. Without entering into the details of the analysis that has been performed, we simply mention that a numerical research of the point of closest approach to the MEMS boundary leads to the results shown by the solid curve in Fig.~\ref{fig3} {\bf (b)}. These reveal that the largest fidelity is achieved at $N=0$ (for an ideally squeezed field). The slower decay of fidelity in the squeezed case might be related to the protected steady state that is typically achieved when this sort of structured baths is considered. 

\subsection{Inclusion of Phase Damping}
\label{phase}
Until now we have only considered dissipation induced by losses in the cavity field due to its coupling with an external bath. This resulted in effective damping terms in the reduced two-qubit master equation. However, spoiling mechanisms intrinsic to the qubit system, such as spontaneous emission and phase-damping, may originate important effects to be taken into account. Typically, these kick in with different time scales and their relative weight is a system-dependent issue. However, a few general considerations can be made here. Spontaneous emission from the excited state of each qubit occurring at rate $\gamma_q\ll{g_j}\ll{\gamma}$ (in order for the bad cavity limit to hold) is formally accounted for by introducing in the two-qubit reduced master equation terms identical to the first line of Eq.~(\ref{reduced}) (for $\nbar=0$ and replacing $g^2_j/\gamma$ with $\gamma_q$). This results in an effective modification of the spontaneous emission rate which becomes $\gamma_q(1+c_j)$ with $c_j=g^2_j/(\gamma_q\gamma)$ the (finite) cooperativity parameter corresponding to qubit $j$. Depending on the relative ratios of the parameters entering the problem, this simply biases the dynamics towards individual spontaneous emission processes of the two qubits, against the effective qubit-qubit interaction, therefore resulting in less entanglement and more mixedness of the stationary state. On the other hand, the inclusion of phase-damping processes in the dynamics encompassed by Eq.~(\ref{reduced}) needs to be commented. 

\begin{figure}[b]
{\bf (a)}\hskip4cm{\bf (b)}
\psfig{figure=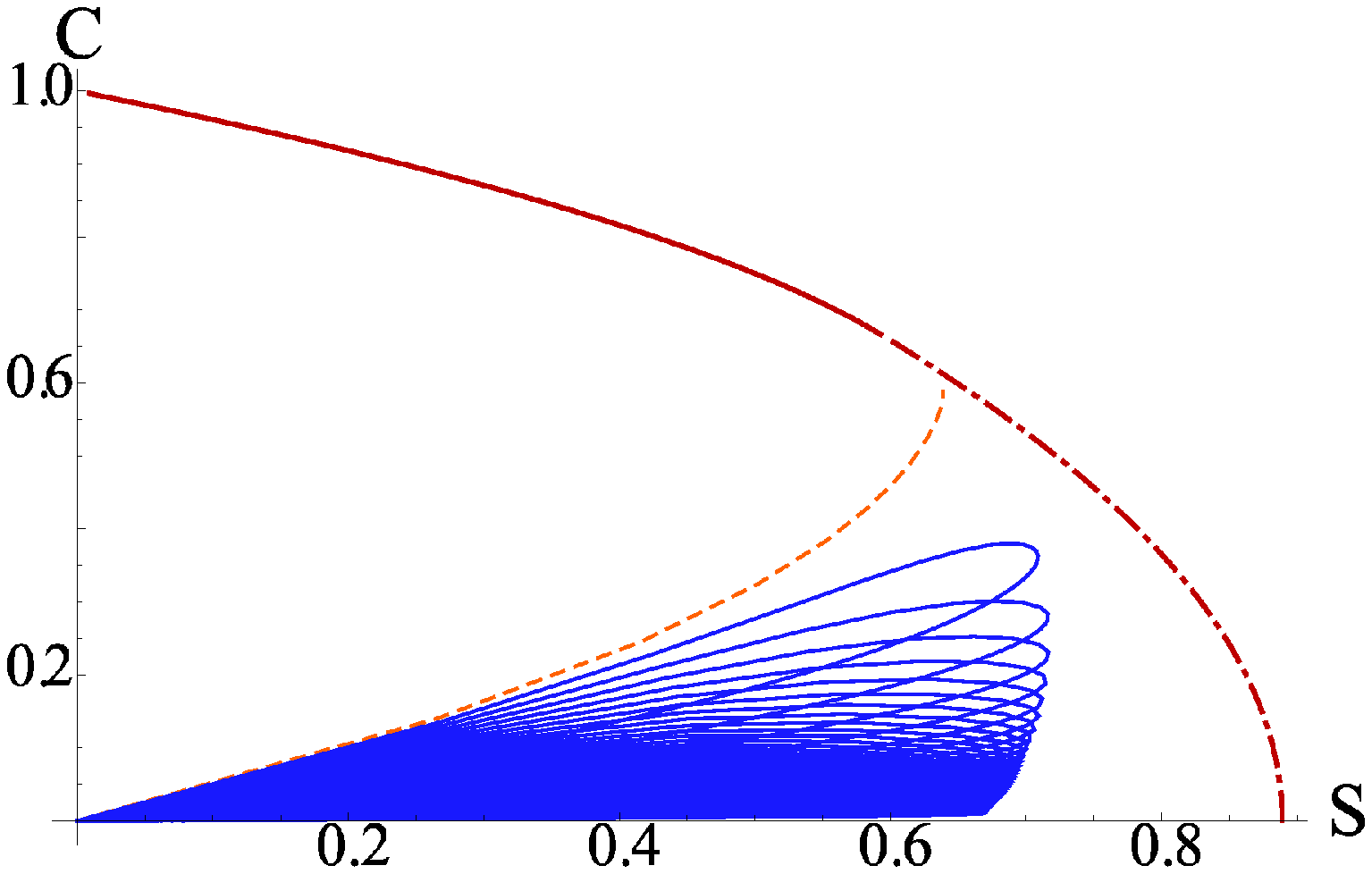,width=4.2cm,height=3.3cm}~~\psfig{figure=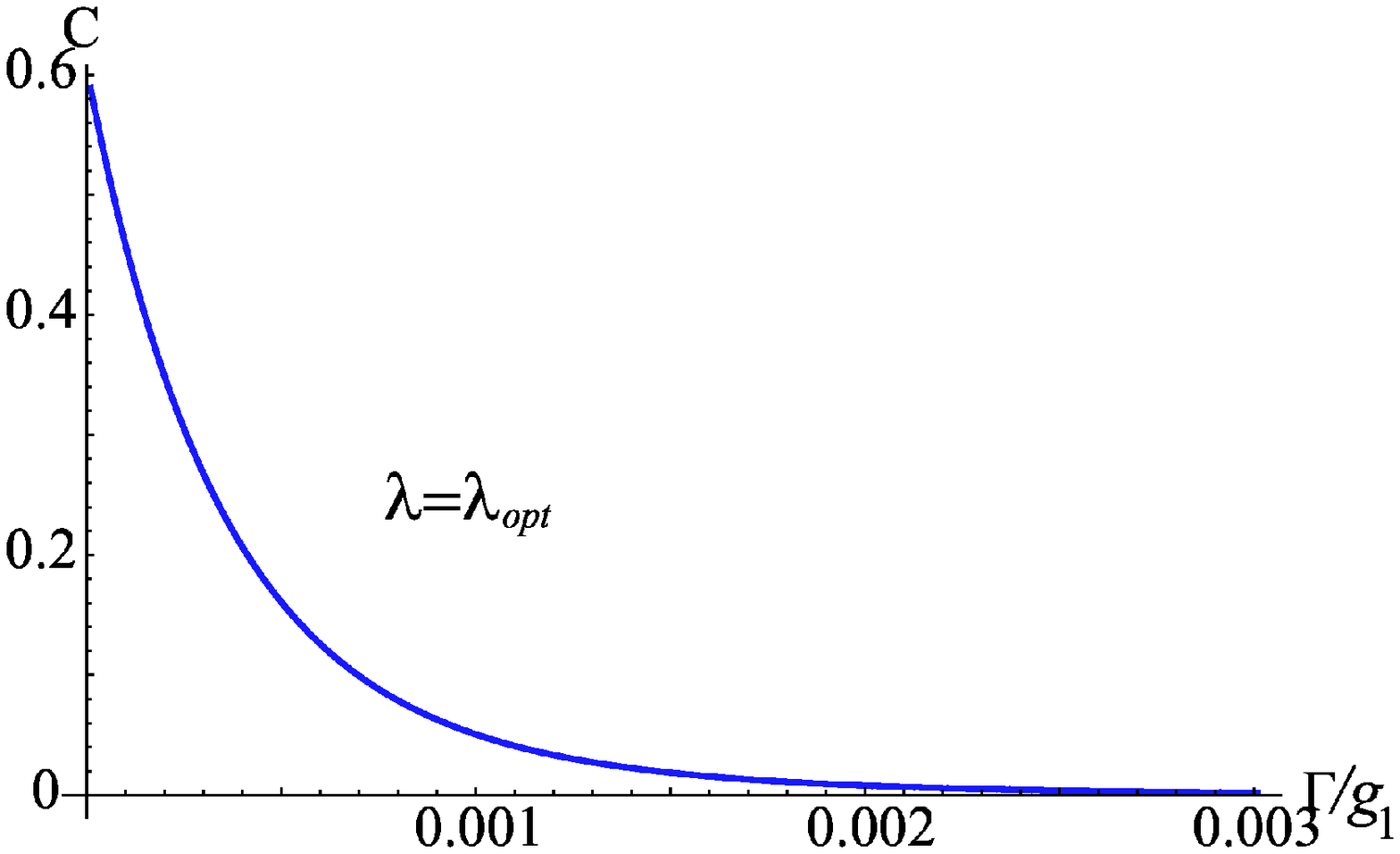,width=4.4cm,height=3.3cm}
\caption{(Color online). {\bf (a)} Effect of phase damping on the open trajectories corresponding to the states with optimal value $\lambda_{opt}$. The dashed curve shows the trajectory corresponding to $\Gamma=0$, while the lower curves show an increasing dimensionless phase-damping rate $\Gamma/g_1$ which goes from $10^{-3}$ to $0.1$ in steps of $10^{-3}$. The lower (dot-dashed) part of the MEMS boundary curve corresponds to MEMS $\rho_2$ affected by independent phase damping channels. {\bf (b)} Decrease in concurrence $C$ against an increasing phase damping rate $\Gamma/g_1$ for the optimal state with $\lambda=\lambda_{opt}$ and $g_1t=100$.}
\label{fig5}
\end{figure}

As the closest state to MEMS occurs for $\nbar = 0$, we only need to consider this case. Phase damping is added by introducing
\begin{equation}
\label{phasedamp}
\hat{\cal L}_{pd}[\varrho(t)]=-\Gamma [\sigma_z , [\sigma_z , \varrho(t) ] ]
\end{equation}
in Eq.~(\ref{ME}). Notice that as $\hat{\cal{L}}_{pd}$ does not involve field operators, it remains unaffected by the procedure used to obtain the reduced qubit dynamics. Therefore, term (\ref{phasedamp}) is straightforwardly added to Eq.~(\ref{reduced}) where we set $\nbar=0$. By means of suitably modified Bloch equations, we are able to quantify the effects that phase damping has on entanglement generation. Fig.~\ref{fig5} shows such effects on our optimal solution. We see that the inclusion of this term has a drastic effect on the state of closest proximity to MEMS. Even for the very small value $\Gamma/g_1 \approx 0.001$, we already see a marked drop in the amount of entanglement our state possesses. 
In fact, Fig.~\ref{fig5} {\bf (b)} shows that by $\Gamma \approx 0.003$ we have lost all the entanglement. Thus, a setting realizing the discussed scheme on effective MEMS's approach should keep phase-damping effects under control in order to mantain the protocol efficient. 

This result is interesting also under another viewpoint. In fact, it can be seen that despite the excellent proximity between $\rho_2$ and our optimal state, the resilience properties of the two states against phase damping processes are quite different: $\rho_2$ as a family of states is incredibly robust to the action of phase-damping channels acting on the qubits. As shown by the dot-dashed part of the boundary curve in Fig.~\ref{fig5} {\bf (a)}, such a channel maps a state belonging to $\rho_2$ into another one within the same family, so that the lower part of the MEMS curves basically folds on itself~\cite{noteMEMS}. On the other hand, the optimal state $\varrho$ is affected in the rather different way addressed above, despite it is less than $1 \%$ away, in terms of state fidelity, from such a MEMS. Such a behavior asks for the design of suitable strategies for the protection of the proprties of $\varrho$, at the optimal point, against processes bringing it away from the MEMS boundary. This is currently the topic of our ongoing investigations. 

\section{Practical consideration}
\label{practical}

Here, we provide some details about an experimental setting we would like to suggest as a potential candidate for the successful implementation of the scheme we have discussed so far. It is somehow obvious that a cavity-QED set-up of two neutral atoms into an optical cavity as well as single-electron-charged quantum dots into a semiconductor microcavity would be well suited for the test of our proposal. However, here we decide to explicitly approach a circuit-QED set-up, which we believe is the most promising scenario for our purposes in virtue of its properties of easy manipulability of stationary qubits and the physical features of the cavity and two-level systems involved.

In this scenario, each qubit is encoded into a standing-still superconducting qubit embodied by a SQUID working in the charge regime at the degeneracy point (to wash away, to first order in the single-Cooper pair charge $2e$, the detrimental effect of low-frequency noise induced by background impurities)~\cite{schon,schoelkopf}. Alternatively, one can use the recently proposed {\it transmon} qubits, a charge-phase qubit that results from a modification of a Cooper-pair box, which achieve umprecedented protection against $1/f$ noise. The transition energy of each superconducting qubit can be adjusted through an external, {\it in situ} magnetic flux that modulates the Josephson energy of the SQUID~\cite{schon} in such a way that the qubit can be easily put in the strong resonant or dispersive regime with the field. This tuning ability is at the basis of the experimentally demonstrated non-demolition measurement of the qubit state through spectroscopic resolution of the field's frequency-pulling effect~\cite{schoelkopf}. The qubits are integrated, via conventional optical lithography, in a full-wave/half-wave on-chip coplanar waveguide split by input/output capacitances at tens of millimeters apart. The capacitors couple the cavity to input/output lines for the injection/leakage of the electromagnetic signal. 

The cavity resonance frequency is in the range of $\sim5$ GHz. With this frequency and an operating temperature $<100$mK, $\nbar$ is as small as $0.06$, which allows for the vacuum-field treatment discussed in detail in this paper.  Multiple-qubits have been experimentally allocated into a single coplanar waveguide, in a way so as to implement two-qubit information transfer via cavity-filed bus~\cite{schoelkopfBUS} and, more recently, a circuit-QED version of the Tavis-Cummings model~\cite{TC}. The biased coupling of qubits $1$ and $2$ can be achieved by embedding them at slight asymmetric locations with respect to a voltage antinode of the sustained field mode. In principle, the stripline is a quasi-unidimensional structure with a very small transversal dimension that reduces the effective volume of the cavity field and enhances the coupling rate with the qubit. This, together with the effective dipole moment of the SQUID qubit ($\sim{2}\times10^{4}ea_0$) gives rise to $g_1/\omega_{f}\simeq{0.2-2\%}$. The energy damping time of the stripline can be as long as $\sim1\mu$s, which in principle allows for a long coherent dynamics within the cavity lifetime (experimental evidences put the qubit damping rate in the range of $2\mu$s) and thus the implementation of the unitary version of the MEMS-approaching scheme. A detailed derivation of the qubit-stripline coupling Hamiltonian and the resulting coupling strength can be found in Paternostro {\it et al.}~\cite{schoelkopf}. As the cavity into which the qubits operate is {\it cut} by interrupting the coplanar waveguide with the input/output capacitors, the resonator quality factor can be electrically tuned from $10^2$ to $10^6$. This would in principle allow the realization of bad-cavity conditions where $g_1\ll{\gamma}$. Therefore, such a set-up is able to probe both the regimes studied here. Finally, we mention that for an individual-qubit it is $\Gamma^{-1}\sim{2}\mu$s, which should put the phase-damping rate in a range of weak effect onto the optimal state properties. Very recently, the ability to perform complete state tomography of two transmon qubits has been experimentally demonstrated~\cite{tomo}, opening up the possibility for preparation, evolution and characterization of the target state.

For the sake of completeness, we just mention that a different scheme for the preparation of MEMS using a pre-arranged off-line entangled resource has been suggested, both in cavity- and circuit-QED, in Ref.~\cite{paternostro}. We refer to that work for an extensive account of the details necessary for such the step.

\section{Concluding Remarks}
\label{remarks}

In this paper we have studied how dissipation can be used in a way so as to engineer a two-qubit state whose properties are fairly close to those of a MEMS. We found that concurrence as large as $C\approx 0.59$ can be set in a state having more than $99\%$ fidelity with $\rho_2$. This was achieved by introducing dissipation in a system than, in the noiseless unitary case, is able to span quite a large portion of the MEMS boundary. Although the noise-affected version of the protocol sees a strong reduction in its ability to produce genuine MEMS, yet a rather considerable possibility of {\it navigation} in the $C-S$ plane is preserved, despite the explicit consideration of dissipation. The relevance of our study  is thus twofold. On one hand, it can be seen as the promising demonstration that a simple protocol for quantum state engineering of entangled mixed state can be designed for dissipation-affected settings. On the other hand, ours is the rigorous and complete quantitative assessment of the ability of a resonant and bias spin-boson coupling to generate MEMS under the effects of relevant sources of noise. As such, we believe our investigation provides valuable and interesting information for the experimental groups interested in quantum state engineering in cavity and circuit-QED, where our study would find a natural and significant implementation. 
%We also examined various different decohering processes on our optimal solution, in the forms of thermal excitations and phase damping. As with any decohering effect our optimal solution looses its entanglement and becomes progressively more mixed.\\
%The thermal effects, as expected, for a large $\nbar $ reduced our state to a maximally mixed one. However in the experimental set-up proposed in \ref{practical} the thermal effects can be well controlled and kept at $\nbar \approx 0.001?$.\\
%An interesting observation was that although our optimal state was very nearly a MEMS it did not exhibit the same robustness to phase damping effects as MEMS. In fact phase damping has a critical effect and effectively destroys all entanglement for a relatively low $\Gamma(\approx 0.003)$. This problem raises the question of how do we gain a high degree of control on phase damping effects?\\
%For completeness we investigated to see if the inclusion of squeezed light could create a larger entanglement, however we found this addition does nothing to improve the entanglement properties of our generated state. 
In perspective, it would be interesting to exploit the apparatus put forward here in order to address whether the system we have considered is able to create {\it boundary entangled mixed states} for more than two qubits, an issue which would require a considerable deal of  theoretical work and represent a stimulating challenge. 
% in quantifying entanglement and mixedness for multipartite systems, and determination of various classes of tripartite MEMS.

\acknowledgments 
We thank Dr. J. F. McCann for discussions. SC thanks DEL for financial support. MP is supported by the UK EPSRC (EP/G004579/1).

\renewcommand{\theequation}{A-\arabic{equation}}
% redefine the command that creates the equation no.
\setcounter{equation}{0}  % reset counter 
\section*{APPENDIX}  % use *-form to suppress numbering
\label{bloch}

As discussed in the main body of the paper, we can solve Eq.~(11) by projecting onto the two-qubit computation basis. In doing so we arrive at a set of differential equations that define the time-behavior of the density matrix. The amount of calculation required is greatly reduced by exploiting the Hermitianity and normalization of the density matrix, thus reducing the set of relevant equations to
\begin{equation}
\label{blochquations}
\begin{split}
\partial_t{\varrho_{0000}}&=\frac{2}{\gamma}[ g_{1}^2 (\nbar + 1) \rho_{1010} + 2 g_{1} g_{2} (\nbar + 1) \rho_{0110}\\
& +g_{2}^2 (\nbar + 1) \rho_{0101} - G^2\nbar \rho_{0000}],
\end{split}
\end{equation}
\begin{equation}
\begin{split}
\partial_t{\varrho_{0001}}&= \frac{1}{\gamma} \{ 2 g_1^2 (\nbar + 1) \rho_{1011} - [g_2^2 (2\nbar +1)+2 g_1^2 \nbar]\rho_{0001}\\
& +2 g_1 g_2 (\nbar + 1)\rho_{0111} - g_1 g_2 (2\nbar + 1) \rho_{0010}\},
\end{split}
\end{equation}
\begin{equation}
\begin{split}
\partial_t{\varrho_{0010}}&= \frac{1}{\gamma} \{2 g_2^2(\nbar + 1)\rho_{0111} -[g_1^2(2\nbar + 1)+2 g_2^2 \nbar]\rho_{0010}\\
& +2 g_1 g_2 (\nbar + 1)\rho_{1011} - g_1 g_2 (2\nbar + 1)\rho_{0001}\}
\end{split}
\end{equation}
\begin{equation}
\begin{split}
\partial_t{\varrho_{0101}}&=\frac{2}{\gamma} \{g_1^2 (\nbar + 1) \rho_{1111}-[g_2^2 (\nbar + 1)+g_1^2 \nbar]\rho_{0101}\\
& +g_2^2 \nbar \rho_{0000}-g_1 g_2 (2\nbar + 1)\rho_{0110}\},
\end{split}
\end{equation}
\begin{equation}
\begin{split}
\partial_t{\varrho_{0110}}&=-\frac{(2\nbar+1)}{\gamma}[g_1 g_2(\rho_{1010}+\rho_{0101})+G^2\rho_{0110}]\\
&+\frac{2g_1 g_2}{\gamma} [(\nbar + 1)\rho_{1111}+\nbar\rho_{0000}],
\end{split}
\end{equation}
\begin{equation}
\begin{split}
\partial_t{\varrho_{0111}}&=-\frac{1}{\gamma}\{[g_1^2 (2\nbar + 1) + 2 g_2^2 (\nbar +1)]\rho_{0111}- 2 g_2^2 \nbar \rho_{0010}\\
&+g_1 g_2 (2\nbar + 1)\rho_{1011} - 2 g_1 g_2 \nbar \rho_{0001}\},
\end{split}
\end{equation}
\begin{equation}
\begin{split}
\partial_t{\varrho_{1010}}&=-\frac{2}{\gamma} \{[g_1^2 (\nbar + 1)+g_2^2 \nbar]\rho_{1010} - g_2^2 (\nbar +1) \rho_{1111}\\
& -g_1^2 \nbar \rho_{0000}+g_1 g_2 (2\nbar + 1)\rho_{0110}\},
\end{split}
\end{equation}
\begin{equation}
\begin{split}
\partial_t{\varrho_{1011}}&=-\frac{1}{\gamma} \{[2 g_1^2 (\nbar + 1)+ g_2^2 (2\nbar +1)]\rho_{1011}-2 g_1^2 \nbar \rho_{0001}\\
& +g_1 g_2 (2\nbar + 1)\rho_{0111}-2 g_1 g_2 \nbar\rho_{0010}\},
\end{split}
\end{equation}
\begin{equation}
\begin{split}
\partial_t{\varrho_{1111}}&=-\frac{2}{\gamma} \{[g_1^2(\nbar +1) + g_2^2(\nbar + 1)]\rho_{1111}-g_1^2 \nbar \rho_{0101}\\
&- g_2^2 \nbar \rho_{1010}-2 g_1 g_2 \nbar \rho_{0110}\}.
\end{split}
\end{equation}
These equations, together with the normalization constraint $\rho_{1111}=1-(\rho_{0000}+\rho_{0101}+\rho_{1010})$ and the de-coupled equation $\partial_t{\rho_{0011}}=-\frac{G^2}{\gamma}(2\nbar+1)\rho_{0011}$, allow for the solution of the dynamical problem discussed throughout the paper.


\begin{thebibliography}{99}
\bibitem{example} For instance, for $r\sim{0.74}$, $\tau\sim1.6$ and $\lambda\sim{0.64}$, the two spectra are given by the four eigenvalues $\{0.74,0.26,0,0\}$.
\bibitem{notatemp} The mean occupation number is related to the equilibrium temperature $T$ of the cavity field and its frequency $\omega_f$ through $\nbar=(e^{\omega_f/K_bT}-1)^{-1}$ with $K_b$ Boltzmann constant.
\bibitem{noteMEMS} This can be easily understood by considering that the two-qubit phase-damping Kraus operators for the sort of channel considered here do not affect the populations of the density matrix but only its coherences, which decay exponentially with $\Gamma$. Therefore, under phase damping, a matrix $\rho_2$ associated to a set value of $r$ is mapped into an isomorphic one, associated with $r'<r$.
\end{thebibliography}
\end{document}